\documentclass[12pt, oneside]{article}
\usepackage{jheppub}
\usepackage{graphicx}
\usepackage{amsmath}								
\usepackage{color}
\usepackage{mathtools}
\usepackage{amssymb}

\usepackage{hyperref}

\hypersetup{
    colorlinks,
    citecolor=blue,
    filecolor=black,
    linkcolor=blue,
    urlcolor=blue
}

\usepackage{overpic}
\usepackage{array}
\usepackage{rotating}
\usepackage{datetime}

\newcommand{\bea}{\begin{eqnarray}}
\newcommand{\eea}{\end{eqnarray}}
\newcommand{\be}{\begin{equation}}
\newcommand{\ee}{\end{equation}}

\begin{document}
\begin{titlepage}
 \thispagestyle{empty}
 \begin{flushright}

 \end{flushright}

 \vspace{30pt}

 \begin{center}
     
  {\fontsize{18}{22} \bf {The geometry of hypermultiplets\footnote{Invited contribution to the book Half a Century of Supergravity, eds. A. Ceresole and G. Dall’Agata.}}}

     \vspace{30pt}
{\fontsize{13}{16}\selectfont {Stefan Vandoren\footnote{Email: s.j.g.vandoren@uu.nl}}} \\[10mm]

{\small\it
Institute for Theoretical Physics  \\
Utrecht University, 3508 TD Utrecht, The Netherlands \\[3mm]}

\vspace{1.5cm}

{\bf Abstract}

\vspace{0.3cm}
   
\end{center}

This is a light and short review on the geometry of hypermultiplets  coupled to supergravity and its appearance in string theory. As is long known, the target space of the hypermultiplet sigma-model is a quaternion-K\"ahler (QK) manifold of negative scalar curvature. We review some aspects of hypermultiplet moduli spaces, but also study aspects of QK-geometries in relation to the swampland, including some concrete examples of swampland QK-manifolds.

\end{titlepage}

\section{Introduction and a bit of my own history}

My PhD advisor at the University of Leuven in Belgium was Antoine Van Proeyen, and later on I did a postdoc at Stony Brook with Martin Ro\v{c}ek and Peter van Nieuwenhuizen. After Stony Brook I got hired as junior faculty  by Bernard de Wit at Utrecht University. It is sort of tempting to think I got raised and educated in a supergravity environment. While partly true, I first started working on hypermultiplet geometry in the context of rigid supersymmetry during my first postdoc in Swansea. Soon after the celebrated Seiberg-Witten paper in '94 on the Coulomb branch of four-dimensional  $N=2$ supersymmetric gauge theories, Seiberg and Witten studied what happened when compactified on a circle down to three dimensions \cite{Seiberg:1996nz}. In three dimensions, vector multiplets can be dualized to hypermultiplets whose scalars, for rigid supersymmetry ($N=2$ in $D=4$, or $N=4$ in $D=3$), parametrize a hyperk\"ahler manifold. I studied the geometry of hyperk\"ahler spaces, and together with some of the experts on instanton calculus in supersymmetric field theory (Nick Dorey, Valentin Khoze and Michael Mattis) and at the time the young and talented PhD student David Tong, we computed some instanton corrections to the Coulomb branch that confirmed \cite{Dorey:1997ij} the predictions from Seiberg and Witten, namely that for gauge group $SU(2)$, the moduli space should be the hyperk\"ahler space constructed by Atiyah and Hitchin \cite{Atiyah:1988jp}. More or less in parallel, I worked with Bernard de Wit et al. on a more general study of how special geometry - the geometry of vector multiplets in $N=2, D=4$ - was inherited by the hyperk\"ahler geometry when compactifying to three dimensions \cite{DeJaegher:1997ka,deWit:1997vg}, a construction more generally called the rigid c-map, see the appendix of \cite{Cecotti:1988qn}. 

From my supergravity mentors, I soon learned the basics of supergravity and its matter couplings in four dimensions with eight supercharges. As discovered in  \cite{Strominger:1990pd,Castellani:1990zd}, the special geometry properties of vector multiplets \cite{deWit:1984rvr} have a more geometric description and realization in string theory in terms of moduli spaces of Calabi-Yau manifolds. The period integrals could then be used to determine quantum corrections to the holomorphic prepotential governing the geometry of $N=2, D=4$ vector multiplets, discussed in other chapters of this book. Since Calabi-Yau compactifications also gave rise to hypermultiplets, and motivated by the early papers \cite{Becker:1995kb,Antoniadis:2003sw} I got triggered  and set myself the goal of determining the quantum corrections to hypermultiplet moduli spaces. Those quantum corrections are governed by the dilaton and determining them would become quite challenging, so it turned out. Over the years 2000-2010, we made quite some progress in a very fruitful and pleasant collaboration with a group of people, most notably with Sergei Alexandrov, Lilia Anguelova, Bernard de Wit, Boris Pioline, Martin Ro\v{c}ek, Frank Saueressig and other PhD's and postdocs at the time. See e.g. \cite{deWit:1999fp,deWit:2001brd,deWit:2001bk,Anguelova:2004sj,Rocek:2005ij,Robles-Llana:2006vct,Robles-Llana:2006hby,Robles-Llana:2007bbv,Alexandrov:2008gh,Alexandrov:2009vj,Alexandrov:2010qdt} for a partial set of references. 

The story of hypermultiplet geometry essentially started back in 1983, with a paper by Bagger and Witten \cite{Bagger:1983tt}. They studied the constraints from $N=2, D=4$ local supersymmetry on the target space of the hypermultiplet sigma model, and it was found that the scalars in the hypermultiplets had to parametrize a quaternion-K\"ahler (QK) manifold. This is in contrast to the case of rigid supersymmetry, where hypermultiplets parametrize hyperk\"aler manifolds. QK geometries will play an important role in this chapter, and so we review them in the next section. Bagger and Witten were not the first to study hypermultiplet couplings to supergravity. Two earlier papers by Zachos \cite{Zachos:1978iw} and Breitenlohner and Sohnius \cite{Breitenlohner:1981sm} considered some special cases of hypermuliplets coupled to supergravity using off-shell methods. For both vector- and hypermultiplet matter couplings, a superconformal tensor calculus was developed in \cite{deWit:1984rvr}. For more on the history of matter couplings in supergravity, we refer to the contribution of Van Proeyen to this book.

The aim of this chapter is to give an overview of the geometry of hypermultiplets, focusing more on the ideas and context, and not so much on techniques and details. I have made an attempt to be somewhat accurate about the history and the original references, but by no means is the list of references complete. There are also other reviews that contain more details and which the interested reader can consult. For instance, I have not reviewed the story of instanton corrections to moduli spaces, as this is done in 2011 in \cite{Alexandrov:2011va} and in 2015 in \cite{Alexandrov:2013yva}. There are also nice reviews on $N=2$ matter-supergravity couplings and their gaugings, elsewhere in this book or in the references \cite{Andrianopoli:1996cm,Trigiante:2016mnt,Lauria:2020rhc}.

\section{Hypermultiplets and quaternion-K\"ahler geometry}

As mentioned in the introduction, it was found by Bagger and Witten that the scalar fields $\phi^i$ in hypermultiplets coupled to four-dimensional supergravity parametrize a quaternion-K\"ahler (QK) manifold \cite{Bagger:1983tt}. Following their conventions, the relevant terms in the Lagrangian are
\begin{equation}
    {\cal L}=-\frac{1}{2\kappa^2}e{\cal R}-g_{ij}(\phi)\partial_\mu\phi^i\partial^\mu\phi^j\ ,
\end{equation}
where ${\cal R}$ is the spacetime curvature, $\kappa^2=8\pi G_N$ is related to Newton's constant and $g$ is the metric on the QK manifold. We have omitted writing down the fermions in the Lagrangian and the supersymmetry transformation rules, from which follows the QK restriction on the target manifold. The reader can find the details in \cite{Bagger:1983tt}.

To the best of my knowledge, this is the first place where QK geometries entered theoretical physics. In mathematics, QK manifolds have a longer history. They are most often defined via the holonomy group. By definition, QK manifolds have (reduced) holonomy contained in the group $Sp(n)\cdot Sp(1)=Sp(n)\times Sp(1)/\mathbb{Z}_2\subset SO(4n)$, where the $\mathbb{Z}_2$ factor is generated by $(-\mathbb{I},-1)$.
As for hyperk\"ahler manifolds (whose holonomy lies in $Sp(n)$), QK manifolds have a real dimension which is a multiple of four, say $4n$.  Since each hypermultiplet contains four real scalars, one needs $n$ hypermultiplets to describe a QK space of dimension $4n$. For $n=1$, since $Sp(1)\cdot Sp(1)\propto SO(4)$, the definition of a QK manifolds must be adjusted and one defines it by requiring the manifold to be Einstein with (anti-)selfdual Weyl curvature. Alternatively, one may define QK's by the existence of having an Sp(1) triplet of almost complex structures. For some math review on QK geometry, see e.g. \cite{SalamonQK}. Despite their name, QK spaces are in general not K\"ahler, as they have no integrable complex structure.
The notion of the quaternionic structure is useful in supergravity, as the three almost complex structures together with the identity generate the supersymmetry transformations.

As a consequence of their holonomy, QK manifolds are Einstein, and therefore they have constant scalar curvature, see e.g. \cite{Salamon}. If the scalar curvature is zero, a QK manifold becomes hyperk\"ahler. Hence one can expect that in a supergravity theory, the scalar curvature is proportional to Newton's constant. A surprising fact is that only QK manfifolds with {\it negative} scalar curvature can be coupled to supergravity. Indeed, Bagger and Witten computed its value to be
\begin{equation}
    R(g)=-8\kappa^2(n^2+2n)\ .
\end{equation}
 Notice that $R/\kappa^2$ is dimensionless and quantized in terms of its quaternionic dimension $n$. In particular this means that the scalar curvature cannot depend on any other parameters in the geometry.

Symmetric QK spaces $G/H$ with positive curvature were classified by Wolf and Alekseevskii \cite{Wolf,Alekseevski}. As shown in the latter reference, all compact homogeneous QK spaces must be symmetric, and in the case when the scalar curvature is positive, they are given by the three infinite series
\begin{eqnarray}
&&\mathbb{H}{\rm{P}}(n)=\frac{Sp(n+1)}{Sp(n)\times Sp(1)}\ , \qquad X(n)=\frac{SU(n+2)}{S(U(n)\times U(2))}\ , \nonumber\\
&&\hspace{2.8cm} Y(n)=\frac{SO(n+4)}{SO(n)\times SO(4)} \ ,
\end{eqnarray}
of dimension $4n$, and the five exceptional cases
\begin{eqnarray}
&&\frac{G_2}{SO(4)}\ , \qquad \frac{F_4}{Sp(3)\times Sp(1)}\ , \qquad \frac{E_6}{SU(6)\times Sp(1)}\ , \nonumber\\
&&\hspace{1cm} \frac{E_7}{Spin(12)\times Sp(1)}\ , \qquad \frac{E_8}{E_7\times Sp(1)}\ ,
\end{eqnarray}
of dimensions 8, 28, 40, 64 and 112 respectively. For low dimensions, there are relations between  these spaces: $Y(1)\cong \mathbb{H}\rm{P}(1)$, and $Y(2)\cong X(2)$. These are the only two complete four-dimensional QK spaces with positive scalar curvature, and correspond to the four-sphere and $\mathbb{CP}^2$:
\begin{equation}
    S^4=\frac{SO(5)}{SO(4)}\ ,\qquad \mathbb{CP}^2=\frac{SU(3)}{S(U(2)\times U(1))}\ .
\end{equation}
Coincidentally, $\mathbb{CP}^2$ is also a K\"ahler manifold, but this is an exception.

As mentioned before, QK spaces with positive curvature cannot be coupled to supergravity and hence do not arise in the low-energy limit of string theory compactifications with eight supercharges. Moreover, QK spaces with positive curvature are quite rigid and there are no known examples 
of non-symmetric QK manifolds with positive scalar curvature. It is furthermore shown that there are only a finite number for each dimension and it is conjectured that all complete QK spaces with positive curvature are symmetric spaces \cite{LeBrun1994}, namely the Wolf spaces.

There exist non-compact versions of the Wolf spaces which have negative scalar curvature and hence are relevant for supergravity applications. For instance, the series 
\begin{eqnarray}
&&\widetilde{\mathbb{H}{\rm{P}}}(n)=\frac{Sp(n,1)}{Sp(n)\times Sp(1)}\ , \qquad \tilde{X}(n)=\frac{SU(n,2)}{S(U(n)\times U(2))}\ , \nonumber\\
&&\hspace{2.8cm} \tilde{Y}(n)=\frac{SO(n,4)}{SO(n)\times SO(4)} \ .
\end{eqnarray}
Similarly, there are non-compact versions of the QK spaces based on the exceptional groups, such as $G_{2(+2)}/SU(2)\times SU(2)$. A classification of QK spaces with negative scalar curvature, relevant for supergravity, does not exist. They are not rigid and can rather easily be deformed preserving the QK properties \cite{LeBrun1991}. In string theory, this will be realized by the perturbative and non-perturbative quantum corrections as we will discuss later. For some earlier studies on the classification of {\it homogeneous} QK spaces with negative curvature, see \cite{Ale75}, \cite{Cecotti:1988ad} and \cite{deWit:1991nm,deWit:1992wf}. Because the $Sp(1)$-curvature does not vanish due to the non-trivial $Sp(1)$ holonomy factor, QK spaces are rather difficult to deal with. This complicates the study of hypermultiplet couplings to supergravity arising from the low energy effective action of superstring compactifications.

Hypermultiplet couplings to supergravity in $D=4,5,6$ with eight supercharges always lead to the same constraints, namely the target space is a negatively curved QK. The Bagger-Witten analysis can easily be adapted to five and six dimensions as well. For early references on the $D=6$ case, see \cite{Nishino:1984gk,Nishino:1986dc}. Upon dimensional reduction from $D=6\to 5 \to 4$, the supersymmetry constraints on the hypermultiplet moduli space remain unchanged. See e.g. \cite{Lauria:2020rhc} for some review about $D=4,5,6$ supergravity with eight supercharges.

In three spacetime dimensions, the story is different, as the supergravity constraints on hypermultiplets require the target space to be a product of two quaternion-K\"ahler manifolds \cite{deWit:1992psp}   (the product of two QK's is generically not a QK). As we will see, this is consistent with compactifying a $D=4$ theory with vector- and hypermultiplets on a circle: the scalar field geometry from the hypers reduces to the same QK in three dimensions, but the vector multiplets produce, after dualizing the vectors into scalars, a second QK manifold. Interestingly, and by contrast, in $D=3$ supergravity with six supercharges (denoted by $N=3$), the target space of scalar fields is again a (single) QK space.

\section{QK's from String Theory}

There are various set-ups from string theory to get to $N=2$ (eight supercharges) supergravity with hypermultiplets from string theory. Many of them involve compactifications of type II superestrings on Calabi-Yau threefolds (CY$_3$), with Hodge numbers $h_{1,1}$ and $h_{1,2}$ governing the K\"ahler and complex structure deformations respectively. We start with these, mention also the connection with heterotic strings, but also discuss some other, more recent, constructions at the end of this section based on asymmetric orbifolds. Before we do so, we mention that in  supergravity coupled to both vector- and hypermultiplets, the total  moduli space consisting of scalar fields from both multiplets  factorize \cite{deWit:1984rvr}, 
\begin{equation}
    {\cal M}={\cal M}_V\times {\cal M}_H\ ,
\end{equation}
where ${\cal M}_V$ is a special K\"ahler geometry determined by a holomorphic prepotential, and ${\cal M}_H$ is a QK space. This means we can study them separately. 
\\

{\bf{QK's from CY$_3$}}\\

Compactifications of eleven-dimensional supergravity or M-theory on a Calabi-Yau threefold (CY$_3$) down to five dimensions gives $h_{1,2}+1$ massless hypermultiplets and $h_{1,1}-1$ massless vector multiplets \cite{Cadavid:1995bk}. After further reduction on a circle $S^1$ to four dimensions, the theory is dual to type IIA string theory compactified on the same CY$_3$, leading to $n_H=h_{1,2}+1$ massless hypermultiplets and $n_V=h_{1,1}$ massless vector multiplets. The compactification of type IIA supergravity was carried out in \cite{Bodner:1990zm}, and it was shown that the dilaton sits in a hypermultiplet. For rigid CY$_3$, having no complex structure deformations, i.e. $h_{1,2}=0$, we obtain the so-called universal hypermultiplet which at string tree level corresponds to the QK manifold
\begin{equation}
\tilde{X}(1)=    \frac{SU(1,2)}{U(2)}\ .
\end{equation}
A parametrization of the QK metric for the universal hypermultiplet is in terms of four real coordinates
\begin{equation}\label{UHM}
    {\rm d}s^2={\rm d}\phi^2+{\rm e}^{-\phi}\Big(({\rm d}\chi)^2+({\rm d}\varphi)^2\Big)+{\rm e}^{-2\phi}\Big({\rm d}\sigma-\chi{\rm d}\varphi\Big)^2\ ,
\end{equation}
where $\phi$ is the dilaton, $\sigma$ the dual scalar to the $D=4$ Kalb-Ramond two-form, and $\chi$ and $\varphi$ two Ramond-Ramond scalars, all obtained from IIA on a rigid CY$_3$.

Similarly, type IIB string theory compactifications on a CY$_3$ yield $n_V=h_{1,2}$ vector multiplets and $n_H=h_{1,1}+1$ hypermultiplets in $D=4$. The compactification of ten-dimensional IIB supergravity on a CY$_3$ was carried out in \cite{Bodner:1989cg,Bohm:1999uk}. We can summarize the multiplet structure of type II theories on (the same) CY$_3$ by
\begin{eqnarray}
    n_V^{A}=h_{1,1}\ ,\qquad n_H^{A}=h_{1,2}+1\ ,\nonumber\\
    n_V^{B}=h_{1,2}\ ,\qquad n_H^{B}=h_{1,1}+1\ ,
    \end{eqnarray}
    where the superscript refers to type IIA or IIB. Notice that switching from IIA to IIB amounts to interchanging the Hodge numbers, which is a manifestation of mirror symmetry of Calabi-Yau threefolds, stating that IIA/CY$_3$=IIB/${\widetilde{\rm{CY}_3}}$, with the tilde denoting the mirror manifold. But as demonstrated in the work of \cite{Cecotti:1988qn} there are also other relations between the vector- and hypermultiplet moduli spaces of IIA and IIB, called the c-map in \cite{Cecotti:1988qn}. In a string theory language, it is simply T-duality between IIA and IIB, both compactified on the same CY$_3$,
    \begin{equation}
        IIA/(CY_3\times S^1_R)=IIB/(CY_3\times S^1_{1/R})\ .
    \end{equation}
    This means that when we reduce the four-dimensional theories further on a circle, one should obtain the IIB moduli spaces from the IIA ones, and vice versa. In particular, the QK moduli space in IIB should follow from the vector multiplet moduli space in IIA, as vector multiplets can be dualized to hypermultiplets once reduced from four to three dimensions. This was understood and worked out by \cite{Cecotti:1988qn,Ferrara:1989ik}, who showed as well how the properties of the (tree-level) QK spaces follow from the special K\"ahler manifolds by dimensional reduction. At tree level in the dilaton, therefore, QK spaces are characterized by a holomorphic function $F(X)$, the prepotential governing the vector multiplet couplings. The classical (in $\alpha'$) contribution to the prepotential in IIA is determined by the triple intersection numbers $d_{ABC}$ of the CY$_3$, and is of the form 
\begin{equation}\label{prepot}
    F(X)=i \frac{d_{ABC}}{6}\frac{X^AX^BX^C}{X^0}\ ,\qquad A,B,C=1,...,n_V=h_{1,1}\ .
\end{equation}
The prepotential is homogeneous of second degree, as required by supersymmetry. One could add the perturbative and worldsheet instanton corrections (in $\alpha'$) to $F$ as well, but we leave them out for simplicity of the presentation. Via the c-map, the full prepotential then also defines a QK moduli space in IIB. A prepotential of the form \eqref{prepot} also has a $D=5$ uplift, where it corresponds to real special geometry \cite{Gunaydin:1983bi}. Given a set of $d_{ABC}$ symbols in \eqref{prepot}, one can work out the corresponding QK spaces. This program was worked out in \cite{deWit:1991nm,deWit:1992wf}, and led to a classification of homogeneous QK spaces in the image of the c-map, including some new examples that were not in the list of Alekseevskii \cite{Ale75}. 

Examples of QK spaces in the image of the c-map with a $D=5$ uplift (hence in terms of $d$-symbols) are the orthogonal Wolf spaces $\tilde{Y}(n)$ with $ n\geq 3$. Also, four of the exceptional Wolf spaces appear as descending from the four magical supergravities in $D=5$, namely the ones based on the $F_4, E_6, E_7$ and $E_8$ groups \cite{Gunaydin:1983rk}. Interestingly, the missing exceptional QK space $G_{2(+2)}/SU(2)\times SU(2)$ can be obtained starting from $D=5$ supergravity with no vector multiplets. This can be realized in M-theory on a CY$_3$ with $h_{1,1}=1$. Any additional hypermultiplets in $D=5$ go along with the ride. After reducing further on a circle to $D=4$, one gets supergravity coupled to one vector multiplet with prepotential $F=i(X^1)^3/X^0$, to which one can apply the c-map. Hence we obtain $G_{2(+2)}/SU(2)\times SU(2)$ as a moduli space in IIB on a CY$_3$ with $h_{1,1}=1$.

For general prepotentials of the type \eqref{prepot}, when the corresponding $d$-symbols are realized as triple intersection numbers of a CY$_3$, then the corresponding QK's are in the string landscape, otherwise they are in the swampland. We return to swampland issues in the next section. 

One can also use the c-map based on different prepotentials. For instance, for quadratic polynomials, 
\begin{equation}
    F=iX^I\eta_{IJ}X^J\ ,\qquad I=0,1,\cdots n\ ,
\end{equation}
where $\eta$ is of signature $(+-\cdots -)$, the c-map gives the unitary Wolf spaces $\tilde{X}(n+1); n\geq 1$ QK spaces \cite{Cecotti:1988qn,Ferrara:1989ik}. Quadratic prepotentials arise as quantum corrections to the cubic prepotential, and are relevant in the neighborhood of conifold-like singularities or any point in the moduli space that picks up a monodromy when circling around it.  
    
The c-map works at tree level in the string coupling $g_s$. Corrections in $\alpha'$ to the prepotential can be carried over to $\alpha'$ corrections to QK spaces via the c-map. It should be stressed that the $\mathbb{H}P(n)$ series are {\it{not}} in the image of the c-map \cite{Cecotti:1988qn,deWit:1991nm}. \\

{\bf{QK's from $K3$}}\\

Hypermuliplets can also be obtained from heterotic string theory. One can start with heterotic string theory compactified on a $K3$ manifold down to six dimensions. In six dimensions with eight supercharges, vector multiplets do not have any scalars. Instead there can be tensor multiplets (each containing one real scalar), and hypermultiplets. The dilaton sits in a tensor multiplet, and so the hypermultiplet moduli space is exact in $g_s$. The total moduli space factorizes again and one obtains a QK from the hypers, just as before. Since heterotic on a $K3$ is dual to F-theory on an elliptically fibered CY$_3$ \cite{Morrison:1996na}, one could argue that again, one obtains QK's from Calabi-Yau threefolds. But the point being here is that QK's do arise in six-dimensional string theories, next to the previous examples of four- and five-dimensional supergravity. 

The total hypermultiplet moduli space of heterotic on K3 is complicated due to the presence of the gauge bundle moduli. See e.g. \cite{Aspinwall:1996mn}. But in the absence of these, by only considering the geometric and $B$-field moduli of the K3, the hypermultiplet moduli space is the QK space\footnote{Here, for simplicity, we are leaving out a further quotient by the discrete duality group.}
\begin{equation}
 {\cal M}_{K3}=   \frac{SO(4,20)}{SO(4)\times SO(20)}\ ,
\end{equation}
which is 80-dimensional. For a more complete description of the hypermultiplet moduli space, it may be useful to go to the F-theory picture, where the number of hypers is given by $n_H=h_{1,2}+1$, and where $h_{1,2}$ correspond to the complex structure deformations of the elliptically fibered CY$_3$. We won't go into more detail here, but still mention that one can study the heterotic hypermultiplet moduli spaces from the known string dualities
\begin{equation}
    Het/(K3\times S^1)=M/CY_3\ ,\qquad Het/(K3\times T^2)=IIA/CY_3\ ,
\end{equation}
where the $CY_3$ is assumed to be $K3$-fibered. The first of these dualities is in five dimensions, the second in four. For some references on heterotic hypermultiplet moduli spaces, see e.g. \cite{Aspinwall:1998bw,Halmagyi:2007wi,Louis:2011aa,Alexandrov:2012pr,Alexandrov:2014jua}.\\

{\bf{QK's from asymmetric orbifolds}}\\

In more recent years, there have been new constructions and examples of hypermultiplet moduli spaces, obtained from freely acting asymmetric orbifolds in four and five dimensions, starting with \cite{Gkountoumis:2023fym,Gkountoumis:2024dwc} and \cite{Baykara:2023plc,Baykara:2024vss}. These are examples of non-geometric compactifications and they have some peculiar features that are not common to the QK's obtained from type II on CY$_3$ or from Heterotic on $K3$.  One is that the dilaton of the type II orbifolds sits in a vector multiplet, and there are no quantum corrections to the hypermultiplet moduli space in either the string coupling $g_s$ or $\alpha'$. Therefore it was possible to determine the exact hypermultiplet moduli space in the examples studied in \cite{Gkountoumis:2024dwc}. These models have typically very few hypermultiplets with target spaces given by the non-compact versions of the Wolf spaces, such as the universal hypermultiplet $\tilde{X}(1)$ and the eight-dimensional symmetric space $\tilde{Y}(2)$. A second peculiar feature is that in some special cases, it is possible to construct orbifolds which have no hypermultiplets at all \cite{Gkountoumis:2023fym,Baykara:2023plc}, see also \cite{Dolivet:2007sz}. This is very different from the geometric compactifications discussed above.

\section{Quaternion-K\"ahler geometry and the swampland}

An natural question to ask in the context of the swampland program \cite{Vafa:2005ui} is whether all QK's can be realized in string theory. As already noticed in \cite{Ooguri:2006in}, the asymptotic region of moduli spaces is conjectured to be a space of negative scalar curvature. Further requirements on the moduli space are completeness and finite volume. For QK spaces coupled to supergravity, the negative curvature is automatically satisfied, even when quantum corrections are included. Still there is a question that remains, namely whether all negatively curved QK's appear in string theory. In this section we consider a few possible examples of swampland QK's and make some general remarks at the end of this section. \\

\textbf{Example 1: Quaternionic hyperbolic spaces $\widetilde{\mathbb{H}{\rm{P}}}(n)$.}\\

Maybe the simplest of all negatively curved QK's are the non-compact versions of the quaternionic projective spaces, namely the quaternionic hyperbolic spaces which we have denoted by 
\begin{equation}
\widetilde{\mathbb{H}{\rm{P}}}(n)=\frac{Sp(n,1)}{Sp(n)\times Sp(1)}\ .
\end{equation}
Their appearance as a target space for hypermultiplets was already shown in \cite{Breitenlohner:1981sm} back in 1981. As mentioned earlier, $\widetilde{\mathbb{H}{\rm{P}}}(n)$ is known not to be in the image of the c-map, and therefore, this candidate hypermultiplet moduli space can not be obtained from a vector multiplet moduli by T-duality.  It will therefore be difficult to realize this QK from a CY$_3$ compactification in string theory. The simplest quaternionic hyperbolic space is the non-compact version of the four-sphere
\begin{equation}
    \frac{Sp(1,1)}{Sp(1)\times Sp(1)}\simeq \frac{SO(1,4)}{SO(4)}\ ,
    \end{equation}
    which is isomorphic to the orthogonal Wolf space $\tilde{Y}(1)$. Viewed as $\tilde{Y}(1)$, it was shown in \cite{Chang:2023pss} that this QK can be obtained from a two-step truncation of the heterotic string on $T^4$, first to four hypermultiplets with moduli space $SO(4,4)/SO(4)\times SO(4)$, and then a further truncation to a single hyper with target $SO(1,4)/SO(4)$. While consistent truncations are valid within supergravity, it remains to be seen if such a truncation has a stringy realization, e.g. via an orbifold of some kind. Even if that would be possible, to realize the quaternionic hyperbolic spaces with $n>1$ seems to be an open problem, and most likely they belong to the swampland.\\

\textbf{Example 2: Homogeneous non-symmetric QK's}\\

The c-map produces a QK manifold of negative scalar curvature out of a special K\"ahler geometry. The latter is the target space for the scalars in the vector multiplet moduli space in $D=4$, and there is a basis in which this is governed by a prepotential $F(X)$, homogeneous of second degree. When $F$ belongs to the string landscape, it follows from the c-map (which in string theory language is nothing but a consequence of T-duality) that the corresponding QK is also in the string landscape. 

Recently, in \cite{Cecotti:2020rjq}, the question was addressed which special K\"ahler geometries are in the swampland. The analysis there shows how sparse the landscape of special geometries is, and most special K\"ahler manifolds actually belong to the swampland. It is natural to conjecture that starting from a prepotential that is in the swampland, the corresponding QK after the c-map is also in the swampland. Concrete examples of cubic prepotentials $F$ that belong to the swampland are discussed in \cite{Cecotti:2020rjq}. When $F$ is purely cubic, the special K\"ahler moduli space should either be locally symmetric for it to have finite volume\footnote{Here, it is important to quotient the moduli space by the discrete group of isometries.}, or otherwise receive quantum corrections. Otherwise it belongs to the swampland. For instance, there are two infinite families of homogeneous but non-symmetric spaces, labeled by integers $(p,q)$, and denoted by $K(p,q)$ and $H(p,q)$ in \cite{Cecotti:1988ad}\footnote{The special cases $K(0,q), H(1,1), H(1,2), H(1,4)$ and $H(1,8)$ do correspond to symmetric cases, and their dual quaternionic spaces are the QK Wolf spaces $\tilde Y(q+4)$ for $K(0,q)$, and those based on four of the exceptional symmetric spaces, respectively.}. The precise form of the prepotential, and other properties of the $K$ and $H$ spaces are discussed in \cite{Cecotti:1988ad}.
The corresponding QK spaces in the image of the c-map are homogeneous and already appear in Alekseevskii's list \cite{Ale75}, they are denoted by $W(p,q)$ and $V(p,q)$ for $K$ and $H$ respectively. The dimension of the QK space $W$ is $4(4+p+q)$. See e.g. \cite{deWit:1991nm} for more information on these and other homogeneous non-symmetric QK spaces. 
\\

\textbf{Example 3: A deformation of the universal hypermultiplet}\\

Let's start with the universal hypermultiplet metric given in \eqref{UHM}. It arises from string theory from compactification of IIA strings on a rigid CY$_3$. Perturbative quantum corrections were studied in \cite{Antoniadis:2003sw}, see also \cite{Anguelova:2004sj,Robles-Llana:2006vct,Alexandrov:2007ec} and it was argued in \cite{Robles-Llana:2006vct} that there is only a one-loop correction to hypermultiplet moduli spaces. Applied to the universal hyper, the one-loop correction can be written as a deformation of the metric \eqref{UHM}, and in the coordinates used in the appendix of \cite{Robles-Llana:2006vct} (with $x^0={\rm e}^{-\phi}$ and a simple rescaling of the other fields), it takes the form
\begin{equation}
    {\rm d}s^2=\frac{1+2c\,{\rm e}^{-\phi}}{1+c\,{\rm e}^{-\phi}}{\rm d}\phi^2+(1+2c\,{\rm e}^{-\phi}){\rm e}^{-\phi}\Big(({\rm d}\chi)^2+({\rm d}\varphi)^2\Big)+\frac{1+c\,{\rm e}^{-\phi}}{1+2c\,{\rm e}^{-\phi}}{\rm e}^{-2\phi}\Big({\rm d}\sigma-\chi{\rm d}\varphi\Big)^2\ .
\end{equation}
Here, the constant $c$ is determined by the Euler number of the CY$_3$
\begin{equation}
    c=-\frac{1}{6\pi}(h_{1,1}-h_{1,2})=-\frac{1}{6\pi}h_{1,1}<0\ ,
    \end{equation}
    where in the last equation we have used that the CY$_3$ is rigid, and therefore the coefficient $c<0$. For $c=0$, one recovers back the classical result \eqref{UHM}. As shown in \cite{Davidse:2005ef,Alexandrov:2009qq}, there is a curvature singularity at ${\rm e}^\phi=-2c$, the location at ${\rm e}^\phi=c$ is a coordinate singularity.  The determinant of the metric is given by
    \begin{equation}
        {\sqrt{g}}={\rm e}^{-2\phi}(1+2c\,{\rm e}^{-\phi})\ ,
    \end{equation}
    which has good fall-off conditions at $\phi\to \infty$, but it shows that for $c<0$, the metric becomes degenerate at ${\rm e}^\phi=-2c$.
The metric is positive definite for ${\rm e}^\phi>-2c$, but as shown in \cite{Alekseevsky:2013nua}, it is incomplete. The singularity at ${\rm e}^\phi=-2c$ should finally be resolved in string theory by including D-brane and NS-fivebrane instanton corrections, and the expectation is that including these non-perturbative correction makes the manifold  complete again with finite volume. For a review on instanton corrected hypermultiplet moduli spaces, see \cite{Alexandrov:2013yva}, and the more recent papers \cite{Alexandrov:2014mfa,Alexandrov:2014rca,Alexandrov:2014sya,Alexandrov:2016tnf,Alexandrov:2017qhn,Alexandrov:2021dyl,Alexandrov:2021shf,Alexandrov:2023hiv}.

Now consider the situation for an arbitrary {\it positive} constant $c$ and notice that the metric is positive definite everywhere. The first fact is that this metric still defines a (non-homogeneous) QK manifold, as follows easily \cite{Anguelova:2004sj,Robles-Llana:2006vct} from the hyperk\"ahler cone description of QK's \cite{Swann1991,Galicki_1992,deWit:1999fp,deWit:2001brd}, see also \cite{Alekseevsky:2013nua} for a more mathematical treatment. But in contrast to the case $c<0$, there is no curvature singularity for $c>0$. In fact, it was shown in the appendix of \cite{Alekseevsky:2013nua}, that the metric for $c>0$ is complete. 

The example with $c>0$ is a bona-fide QK manifold which is complete and negatively curved. Yet, is seems to belong to the swampland because string theory based on a rigid CY$_3$ has a positive Euler number and hence gives $c<0$. Most likely, there exist no quotient by a discrete isometry group that makes the volume finite, and hence one of the moduli space swampland conjectures \cite{Ooguri:2006in} is violated. 
Perhaps there are other stringy constructions where this geometry might arise, possibly with additional quantum corrections, but to the best of my knowledge, the only other example where the QK space $SU(2,1)/U(2)$ appears in string theory is in the orbifold construction of \cite{Gkountoumis:2024dwc}. But in that case, the classical hypermultiplet moduli space is exact and there are no quantum corrections because the dilaton sits in a vector multiplet.\\

\textbf{Some general remarks}\\

We end this section with some general remarks concerning the swampland. There is quite some mathematics literature that might be helpful understanding the QK landscape. Two main properties, besides the negative curvature (automatically satisfied for hypermultiplets), are completeness and finite volume. \\

\begin{itemize}
    
\item{\textbf{Completeness}}

About completeness of QK manifolds, it is shown in \cite{Cortes:2011aj,Cortes:2013gda} that completeness is preserved under the c-map. So whenever the vector multiplet moduli space is complete, so is the hypermultiplet moduli space after the c-map. There is a similar statement about the so-called r-map, which maps $D=5$ real special geometry to $D=4$ special K\"ahler geometry upon dimensional reduction. Homogeneous QK spaces obtained in this way are automatically complete, but generically they receive quantum corrections in the dilaton which 
spoils the completeness. An exception are the hypermultiplet moduli spaces constructed in \cite{Gkountoumis:2024dwc}, which are classically exact and as freely acting orbifolds have a higher $N>2$ supersymmetric origin. 

An interesting example of this completeness-construction is the complete, but inhomogeneous special K\"ahler space governed by the prepotential which is a deformation of the $STU$-model and which arises from M-theory on CY$_3\times S^1$ or the dual heterotic model:
\begin{equation}\label{prepot2}
    F=i\Big(\frac{X^1X^2X^3}{X^0}+\frac{1}{3}\frac{(X^3)^3}{X^0}\Big)\ .
\end{equation}
This prepotential is still cubic of the form \eqref{prepot}, but the geometry is inhomogeneous due to second term. As shown in \cite{Cortes:2011aj}, it leads to a complete and inhomogeneous QK manifold of dimension 16, which is a deformation of the orthogonal Wolf space $\tilde{Y}(4)$. 

For more results on completeness theorems for quaternion-K\"ahler spaces, including deformations of general c-map spaces, see \cite{Cortes:2016wjn,Cortes:2017yrp}.

\item{\textbf{Finite volume}}

An equally important swampland constraint is the finiteness of the volume of the moduli space. In fact, the special K\"ahler manifold based on \eqref{prepot2} is complete, but since it is not a Hermitian symmetric space, there is no discrete subgroup one can quotient by such that the resulting manifold has finite volume \cite{Cecotti:2020rjq}. This would mean that \eqref{prepot2} belongs to the swampland, and hence its QK-partner in the image of the c-map. In fact, there is a theorem proven directly for QK manifolds, saying that the homogeneous QK's only admit quotients of finite volume if and only if it is symmetric \cite{AC1999}. 

A way out for these spaces to belong to the string landscape, is when there are quantum corrections. These can be worldsheet instanton corrections to the prepotential (which we know to be the case in type IIA on elliptically fibered CY$_3$), or perturbative and non-perturbative corrections to the QK metric, as mentioned before. 

Very recently, an interesting paper appeared \cite{Cortes:2023oje} that discussed finiteness of the volume of D(-1)-D1 instanton-corrected hypermultiplet moduli spaces in IIB \cite{Robles-Llana:2006hby,Alexandrov:2009qq,Alexandrov:2012bu} compactified on a CY$_3$. This moduli space has an exact SL(2,$\mathbb{Z}$) symmetry that, when quotiented by, can be shown to have finite volume \cite{Cortes:2023oje}. The resulting QK-space is however not complete as expected, since it does not contain all the instanton corrections, such as those coming from wrapped Euclidean NS-fivebranes over the CY$_3$. Clearly, all these aspects require further study.

\end{itemize}

\section*{Acknowledgments}

 I thank all my collaborators over the years who have worked on this topic, many of whom are mentioned in the introduction. Furthermore, I thank Sergei Alexandrov, Boris Pioline and Antoine Van Proeyen for comments and feedback on this manuscript. I also thank the editors A. Ceresole and G. Dall’Agata for inviting me to write this chapter.

\bibliographystyle{JHEP}
\bibliography{ref}

\end{document}